\documentclass{fizik}
\usepackage{hyperref}
\hypersetup{
colorlinks=true,
urlcolor=blue,
citecolor=blue}
\usepackage[all]{xy,xypic}

\usepackage{amsfonts,amssymb,amsmath,amsgen,amsopn,amsbsy,theorem,graphicx,epsfig}
\usepackage{eufrak,amscd,bezier,latexsym,mathrsfs,enumerate,multirow}
\usepackage[utf8]{inputenc}\usepackage[english]{babel}
\usepackage[dvipsnames]{xcolor}
\usepackage{tikz}
\usetikzlibrary{arrows.meta} 
\usepackage{array}

\usepackage{amsmath}
\usepackage{xparse}
\usepackage{physics}
\usepackage{graphicx}
\usepackage{subcaption} 
\usepackage[pagewise]{lineno}
\usetikzlibrary{matrix}

\yil{}
\vol{}
\fpage{}
\lpage{}
\doi{}

\title{Coupling a discrete state to a quasi-continuum: A model 
quantum mechanical system that interpolates between Rabi 
oscillations and decay-revival dynamics}

\author[İŞGÖRÜR, CEVHEROĞLU, and ÖZAKIN]{
\textbf{Enes Kutay İŞGÖRÜR$^{1}$\href{https://orcid.org/0009-0000-8157-6201}{}, 
Osman CEVHEROĞLU$^{2}$\href{https://orcid.org/0009-0004-4572-5379}{}, Arkadaş ÖZAKIN$^{2}$\thanks{arkadas.ozakin@bogazici.edu.tr}~\href{https://orcid.org/0000-0002-2318-9227}{}}\\
$^{1}$Department of Physics, Faculty of Engineering, Free University of Berlin, Berlin, Germany\\
$^{2}$Department of Physics, Faculty of Arts and Sciences, Boğaziçi 
University, İstanbul, Turkey

\\ [1.8em]

}

\newcommand{\bc}{\begin{center}}
\newcommand{\ec}{\end{center}}

\numberwithin{equation}{section}

\renewcommand{\phi}{\varphi}

\begin{document}

\maketitle

\begin{abstract}We formulate a quantum mechanical system 
consisting of a single discrete state coupled to an 
infinite ladder of equally-spaced states, 
the coupling between the two being
given by a Lorentzian profile. Various limits of this system
correspond to well-known models from quantum optics, namely,
the narrow resonance limit gives the Rabi system, 
the wide resonance limit gives the Bixon-Jortner
system, the wide resonance, true continuum limit
gives the Wigner-Weisskopf system, and the fixed resonance, true continuum limit
gives a system that is typically studied by methods developed by Fano. We give a semi-analytical
solution of the eigenvalue problem by reducing it
to a transcendental equation, and demonstrate the aforementioned
limiting behaviors. We then study the dynamics of the initial discrete
state numerically, and show that it gives a wide range of
behaviors in various limiting cases as predicted by our asymptotic 
theory including exponential decay, revivals, Rabi oscillations, and 
damped oscillations. The ability of this system to interpolate 
between such a rich set of behaviors and existing model systems, and the accessibility of a semi-analytical solution, make it a useful model system in quantum optics and related 
fields.

\end{abstract}

\section{Introduction}

The two-state system in quantum mechanics is a valuable model whose 
exact solution is a source of intuition for a wide 
range of systems in quantum optics, atomic and molecular physics, 
and even particle physics.
The general two-dimensional Hamiltonian is easily diagonalized, giving the dynamics of the system in terms of Rabi oscillations.
See \cite{allen2012optical} 
for a detailed review of the use of the two-state approximation in the context of atomic physics. Another model system of comparable utility is the so-called Wigner--Weisskopf
model where the Hilbert space has a basis consisting of a 
single discrete state together with a (1-dimensional) continuum of 
states, with a ``flat'' coupling between these two pieces.
This system can be solved analytically as well, 
resulting in an instability for the discrete state,
whose dynamics being given by an exponential, irreversible decay, 
with a decay constant which can be obtained exactly from the Fermi golden rule.
This decay behavior, as well, comes up in a wide variety of systems,
including examples from atomic physics and particle physics.
The exact solution in this case was first derived by Wigner and Weisskopf \cite{weisskopf1930berechnung}, and an elementary derivation is given in \cite{cohen1986quantum}. See \cite{barnett2002methods}
for a discussion in the context of quantum optics.


These two types of behavior, namely, coherent oscillations of the Rabi 
system and the irreversible decay of the Wigner-Weisskopf system, 
are two archetypes of quantum evolution, both providing
useful approximations in a wide variety of settings.

There are two important generalizations of the Wigner-Weisskopf system which can also be solved analytically or semi-analytically. 
One generalization, 
called the Bixon-Jortner system \cite{bixon1968intramolecular}, 
involves replacing the continuum with a quasi-continuum 
that consists of an infinite set (a ``ladder'') of states with a small energy spacing, 
while still keeping a flat (constant) coupling between the discrete state and the quasi-continuum. 
In the limit of zero ladder spacing, the behavior of this system converges to
that of the Wigner-Weisskopf system, but at finite spacing, the system displays departures
from exponential decay such as revivals after an initial decay. Quantum revivals provide
a very interesting testground for both theoretical and experimental investigations
of quantum behavior (see \cite{bluhm1995evolution} and \cite{narozhny1981coherence}  for some examples),
and the Bixon-Jortner system is one of the simplest settings
where one can study such dynamics analytically.
The
original context where the system was first described
was that of quantum chemistry and the problem of
``intramolecular radiationless transitions'' \cite{bixon1968intramolecular},
and thus the BJ system has direct practical use, as well.



The other generalization of the Wigner-Weisskopf system involves
keeping the continuum as a real continuum, but choosing a varying
coupling coefficient between the discrete state and the continuum, 
in the sense that the coupling depends on the unperturbed energy eigenvalues of the continuum.  Perhaps the most
important analytically solvable case consists of using a Lorentzian resonance profile for the coupling as a function of the continuum energy. This generalization
results in more complicated (and realistic) behavior, including an analogue of the 
Lamb shift in eigenvalues, in addition to decay dynamics. 
In the limit of a wide resonance, one recovers the Weisskopf-Wigner flat coupling behavior, and in the opposite limit of a narrow coupling, one effectively ends up with a coupling to a single
state of the continuum, which gives Rabi oscillations for the initial discrete state. See
\cite{barnett2002methods65} 
for a derivation using techniques developed by Fano \cite{fano1961effects} and a detailed discussion.

The two generalizations of the Weisskopf-Wigner system mentioned,
namely, the generalization to
a non-flat coupling and the generalization to a quasi-continuum,
each allow us to cover a richer class of behaviors. For the 
quasi-continuum, we get revivals in addition to the 
decay, and for the non-constant (Lorentzian) coupling, we get 
an analogue of the Lamb shift, and also cover
a range of behaviors interpolating between Rabi oscillations and
Weisskopf-Wigner decay.

In this paper, we define, and semi-analytically solve,
a simultaneous generalization in both of these directions.
Namely, we consider a coupling of a discrete state
to a \textbf{quasi-continuum} as in the Bixon-Jortner system,
but we take the \textbf{coupling to be non-constant}, as in the Lorentzian
generalization of the Weisskopf-Wigner system. This doubly-generalized system 
interpolates between all four systems described above. By 
investigating the effects of the tuning parameters,
it is possible
to obtain the behaviors of the Rabi system, the Bixon-Jortner system, the Wigner-Weisskopf system,
and the Lorentzian continuum in approppriate limits. A generic choice of parameters results in a general, much richer range of behaviors.

The rest of the paper is organized as follows. 
In Section \ref{review}, we give a review of the Bixon-Jortner,
Rabi, Wigner-Weisskopf, and the Fano/Lorentzian systems. In Section \ref{3}, we define our generalization of the BJ system. In Section \ref{4}, we derive our solution, and
in Section \ref{5}, we investigate the properties of the
solution by considering various limits. In Section \ref{6}, we describe the dynamics of the system when it starts in the 
discrete state, showing the limiting cases of Rabi oscillations,
Fermi golden rule decay, Bixon-Jortner decay-revival dynamics, and 
the more general behaviors interpolating between these limits.
Finally, in Section \ref{conc}, we discuss our results.
Throughout the paper, we work in units where $\hbar=1$.

\begin{table}[h!]
\centering
\renewcommand{\arraystretch}{3}
\setlength{\tabcolsep}{16pt}

\begin{tabular}{c|c|c}
    & \textbf{Constant Coupling} & \textbf{Lorentzian Coupling} \\
    \hline
    \textbf{Continuum} & Weisskopf--Wigner & ``Fano'' \\
    \hline
    \textbf{Discrete} & Bixon--Jortner & ? \\
\end{tabular}
\caption{The model systems of quantum optics and our contribution. 
The generalized, Lorentzian Bixon-Jortner system we study in this
paper fills in the question mark.}
\label{tab:coupling_models}
\end{table}

\section{Review of relevant systems}\label{review}

In this section, we give a quick review of the Bixon-Jortner, Rabi,
and Weisskopf-Wigner
systems in order to motivate the generalization we
will be pursuing, and to set the notation for the limiting cases
we will derive later. For the Bixon-Jortner system,
we mostly use the notation of \cite{tannoudji1992atom}. 

\subsection{The Bixon-Jortner system}\label{2.1}

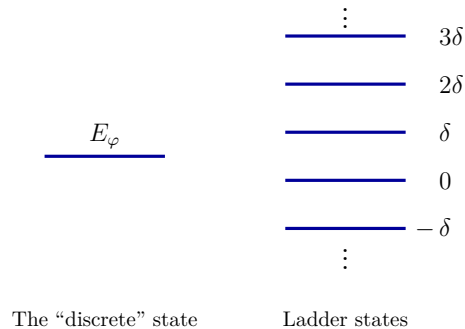
\begin{figure}[h!]\label{fig:bj-schematic}
    \centering
    \begin{tikzpicture}[
        scale=0.8, transform shape, 
        level/.style={draw=blue!60!black, very thick},
        desc_label/.style={text width=3.5cm, align=center},
        elabel/.style={anchor=west, font=\large}
    ]
        \def\spacing{0.8}

        \foreach \y in {0, 1, 2, 3, 4} {
            \pgfmathtruncatemacro{\n}{\y-1} 
            \draw[level] (1, \y*\spacing) -- (3, \y*\spacing); 
            \node[elabel] at (3.4, \y*\spacing) { 
                \ifnum\n=0 $0$\fi
                \ifnum\n=1 $\delta$\fi
                \ifnum\n>1 $\n\delta$\fi
                \ifnum\n=-1 \makebox[0pt][r]{$-$}$\delta$\fi
                \ifnum\n<-1 \pgfmathtruncatemacro{\nabs}{abs(\n)}\makebox[0pt][r]{$-$}$\nabs\delta$\fi
            };
        }
        \node[font=\Large] at (2, 4.5*\spacing) {$\vdots$}; 
        \node[font=\Large] at (2, -0.5*\spacing) {$\vdots$}; 
        \node[desc_label] at (2, -1.5) {Ladder states}; 

        \draw[level] (-3, 1.5*\spacing) -- (-1, 1.5*\spacing) node[midway, above, font=\large] {$E_{\varphi}$};
        \node[desc_label] at (-2, -1.5) {The ``discrete'' state};

    \end{tikzpicture}
    \caption{Schematic representation of the unperturbed eigenvalues of the Bixon-Jortner system. The perturbation $V$ introduces a constant coupling between the discrete state and the ladder.}
    \label{fig:energy_levels}
\end{figure}

The Bixon-Jortner system consists of a two-piece Hamiltonian
$H = H_0 + V$ defined on a Hilbert space spanned by the so-called ``discrete'' state $|\varphi\rangle$ and a 
``ladder'' (or quasi-continuum) of states $|n\rangle$, where $n = 0, \pm 1, 
\pm 2, \ldots$.
These basis vectors are eigenstates of $H_0$:
\begin{align}
    H_0 |\varphi\rangle &= E_{\varphi}|\varphi\rangle\\
    H_0 |n\rangle &=n \delta |n\rangle\,.
\end{align}
See Figure \ref{fig:energy_levels} for a visual representation. The $V$ piece of the Hamiltonian couples the discrete state to the quasi-continuum
with a constant coupling coefficient $v$,\footnote{We will assume $v$ 
is real, a complex $v$ simply adds a phase to the eigenvectors.}
\begin{align}
    \langle n| V |\varphi\rangle &= v = \langle \varphi| V |n\rangle\\
    \langle n| V |n\rangle &= 0 = \langle \varphi |V|\varphi\rangle.
\end{align}
The eigenvalues $E_{\mu}$ and eigenvectors $|\psi_{\mu}\rangle$ of the full
Hamiltonian $H = H_0 + V$ can be found by computing the inner product
of the Schrödinger equation 
$H|\psi_{\mu}\rangle = E_{\mu}|\psi_{\mu}\rangle$
with $\langle\varphi|$ and $\langle n|$, respectively,
\begin{align}
   \langle \varphi|(H_0 + V)|\psi_{\mu}\rangle &= \langle \varphi|\psi_{\mu}\rangle E_{\mu}\\
   \langle n|(H_0 + V)|\psi_{\mu}\rangle &= \langle n |\psi_{\mu}\rangle
   E_{\mu}\,.
\end{align}
By inserting a resolution of identity, these give rise to
a transcendental equation for the eigenvalues $E_{\mu}$,
\begin{equation}\label{BJ_eigval}
    \frac{\pi v^2}{\delta} \cot\left(\frac{\pi E_{\mu}}{\delta}\right) = E_{\mu}\,.
\end{equation}
See \cite{tannoudji1992atom} 
for a derivation and Figure \ref{fig:a_zero_limit_eigval} for a visual representation of the 
solutions.
Making use of the normalization constraint,
\begin{equation}
       |\langle\varphi|\psi_{\mu}\rangle|^2 + \sum_{n=-\infty}^{\infty}|\langle n|\psi_{\mu}\rangle|^2 = 1\,,
\end{equation}
and a choice of phase, 
the components $\langle \varphi|\psi_{\mu}\rangle$
and $\langle k|\psi_{\mu}\rangle$ of the eigenvectors $|\psi_{\mu}\rangle$ 
can be obtained 
in terms of the energy eigenvalues as,
\begin{align}
    \langle \varphi|\psi_{\mu}\rangle &= \frac{v}{\left(v^2 + \left(\frac{\Gamma}{2}\right)^2 + E_{
    \mu}^2\right)^{1/2}} \label{bj-eigvecs-1}\\
    \langle k|\psi_{\mu}\rangle &= \left(\frac{v}{E_{\mu}-k\delta}\right)\langle \varphi|\psi_{\mu}\rangle\,,\label{bj-eigvecs-2}
\end{align}
where $\Gamma = \frac{2\pi v^2}{\delta}$.

The dynamics $|\psi(t)\rangle$ of the initial state
$|\psi(0)\rangle = |\varphi\rangle $ can be found by  
using the eigenvector decomposition of $|\varphi\rangle$. While
the general behavior is rather complex (involving both
decays and revivals, as depicted in Figure \ref{fig:decay-plots2}
), in the limit of small spacing $\delta\to 0$ (taken
while keeping $\Gamma$ constant), one gets an exponential decay given by,
\begin{equation}
    |\langle\varphi|\psi(t)\rangle|^2 = e^{-\Gamma t}\,.
\end{equation}
This is exactly the behavior one would get by
using the Fermi golden rule using the density of states $\rho = 1/\delta$. 
\subsection{The Rabi system}
The Rabi system, or the two-state system is the simplest nontrivial quantum mechcanical system.
We write its general (time-independent) Hamiltonian as,
\begin{equation}
    H = \begin{bmatrix} E_1 & v^* \\ v & E_2 \end{bmatrix} = \alpha_0 I + \mathbf{\alpha}\cdot \mathbf{\sigma}\,.
\end{equation}
where $\mathbf{\sigma} = (\sigma_x, \sigma_y, \sigma_z)$ denotes the triple of Pauli matrices, 
and the real constants $\alpha_i$, $i=0, 1, 2, 3$ are given as, 
\begin{align}
     \alpha_0 = \frac{E_1 + E_2}{2}\,, \quad{}
    \alpha_3  = \frac{E_1-E_2}{2}\,, \quad{}
    \alpha_1  + i \alpha_2 &= v\,.
\end{align}
The two eigenvalues of this Hamiltonian are given as,
\begin{align}
    E_{\pm} &= \alpha_0 \pm \sqrt{\alpha_1^2 + \alpha_2^2 + \alpha_3^2}
    = \left(\frac{E_1+E_2}{2}\right) \pm \sqrt{\left(\frac{E_1-E_2}{2}\right)^2 + |v|^2}\,.
\end{align}
The corresponding eigenvectors are,
\begin{align}\label{eqn:rabi-eigvecs}
    |\psi_{+}\rangle = \begin{bmatrix}\cos{(\theta/2)}e^{-i\phi/2}\\ \sin{(\theta/2)e^{i\phi/2}}\end{bmatrix}\,,\qquad{}
    |\psi_{-}\rangle = \begin{bmatrix}-\sin{(\theta/2)}e^{-i\phi/2}\\ \cos{(\theta/2)e^{i\phi/2}}\end{bmatrix}\,,
\end{align}
where,
\begin{align}
    \tan{\theta}=\frac{2 \abs{W}}{E_1-E_2}\,,\quad{}
    v = \abs{v} e^{i\phi}\,.
\end{align}
See \cite{cohen1986quantum1}. 
For later reference, we also derive formulas for the special case of  $E_2=0$. In this
case, the Hamiltonian is,
\begin{align}\label{eq:rabi-h-e2-zero}
    H = \begin{bmatrix}
        E_1 & v^* \\ v & 0
    \end{bmatrix}\,,
\end{align}
and the eigenvalues become,
\begin{align}\label{Rabi-eigenval}
    E_\pm=\frac{E_1}{2}\pm\sqrt{\frac{E_1^2}{4}+\abs{v}^2} \,.
\end{align}
Noting that $|E_-| = -E_-$ (since $E_- \le 0$), 
one gets,
\begin{align}
\cos{\theta/2}  &=\frac{E_+}{\sqrt{E_+^2+|v|^2}}=
\frac{|W|}{\sqrt{E_-^2+|v|^2}}\\
    \sin{\theta/2}&=\frac{|W|}{\sqrt{E_+^2+|v|^2}} = \frac{-E_-}{\sqrt{E_-^2+|v|^2}}\,.
\end{align}
These formulas allow us to combine the two eigenvector formulas \eqref{eqn:rabi-eigvecs}
into a single formula in terms of the corresponding eigenvalues:
\begin{align}\label{eq:rabi-eigvecs-ours}
    |\psi_\pm\rangle=\frac{1}{\sqrt{E_{\pm}^2+|v|^2}}\begin{bmatrix}
    E_{\pm}e^{-i\phi/2}\\ 
    |v|e^{i\phi/2}\end{bmatrix}\,.
\end{align}
For the case of a real $v$ (so that $\phi=0$), this becomes,
\begin{align}\label{eq:rabi-eigvecs-ours-real-v}
    |\psi_\pm\rangle=\frac{1}{\sqrt{E_{\pm}^2+v^2}}\begin{bmatrix}
    E_{\pm}\\ 
    v\end{bmatrix}\,.
\end{align}
Note that these formulas are rather non-standard in that they give the eigenvectors directly in terms of the 
corresponding eigenvalues. 

\subsection{The Weisskopf-Wigner and Fano solutions} \label{2.3}

The case of a real continuum of states coupled to a single, ``discrete'' state 
can be obtained as the limit of small spacing $\delta\to 0$ of a quasi-continuum like the 
Bixon-Jortner system. One can solve
the true continuum case more directly using a
technique by Fano \cite{fano1961effects}. 
Here we give a quick summary of a specific version that will be relevant to our system.

We assume the Hilbert space of the system is spanned by the discrete state $|\varphi\rangle$
and a continuum $|E\rangle$ of states labeled by the unperturbed energy eigenvalue $E\in \mathbb{R}$, having Dirac delta normalization, 
$\langle E|E'\rangle = \delta(E-E')$. 
We assume the total Hamiltonian is given as,
\begin{align}\label{hamiltonian-derssed}
H=E_\varphi|\varphi\rangle\langle\varphi|+\int E|E\rangle\langle E|dE+\int v(E)\left[|E\rangle\langle \varphi|e^{-i\varphi(E)}+|\varphi\rangle\langle E|e^{i\varphi(E)}\right]dE\,.
\end{align}
We denote the eigenvectors of the 
total (perturbed) Hamiltonian as $|\psi(E')\rangle$, where $E'$ denotes 
the eigenvalue. The eigenvectors can be expanded in terms of the 
``bare'' eigenstates $|E\rangle$ as,
\begin{align}\label{bare-state-basis-eq}
|\psi(E')\rangle=\alpha(E')|\varphi\rangle+\int\beta(E',E)|E\rangle dE\,.
\end{align}
Substituting \eqref{bare-state-basis-eq} and \eqref{hamiltonian-derssed} in 
the eigenvalue equation,
\begin{align}
H|\psi(E')\rangle&=E'|\psi(E')\rangle\,.
\end{align}
and taking the inner product with $|\varphi\rangle$ and $|E\rangle$, 
respectively, one gets a set of equations that need to be solved for
the coefficients $\alpha(E')$ and $\beta(E', E)$. This procedure gives
(see \cite{barnett2002methods} for details),
\begin{align}\label{alphasquared}
    |\alpha(E')|^2&=\frac{W^2(E')}{[E'-F(E')-E_\varphi]^2+\pi^2W^4(E')}\,.
\end{align}
where $F(E')$ is given as the principal value integral,
\begin{align}
    F(E') &= \mathbb{P}\int\frac{W^2(E)}{E'-E}dE\,,
\end{align}
and $z(E')$ is given as,
\begin{align}
    z(E') &=\frac{E'-F(E')-E_\varphi}{W^2(E')}\,.
\end{align}
One then obtains $\beta$ directly in terms of $\alpha$.
Specializing to the case of $E_{\varphi}=0$, 
the Wigner-Weisskopf case is given by a flat (constant) coupling $W(E)=W$.
For this case one has $F(E')=0$, and with a choice of phase, $\alpha$ becomes,
\begin{align}
    \alpha(E')=\frac{W}{\sqrt{(E')^2 +\pi^2 W^4}}\,.
\end{align}
If instead of a constant coupling, one uses a ``resonant coupling'' $W^2(E')=\frac{W^2\gamma}{\pi((E')^2+\gamma^2)}$ where $W$ and $\gamma$ are constants,\footnote{Here we follow the
convention of \cite{barnett2002methods}, which makes the units 
of the constant $W$ different from the units of the $W$ in the Wigner-Weisskopf case.}  $F(E')$
and $|\alpha|^2$ become,
\begin{align}
    F(E')&=\frac{W^2E'}{(E')^2+\gamma^2}\\
    |\alpha(E')|^2&=\frac{W^2\gamma}{\pi}\frac{1}{\bigl[(E')^2-W^2\bigr]^2+\gamma^2(E')^2}\label{alphaabs}\,.
\end{align}
See \cite{barnett2002methods} for details.
If the system is started in the discrete state $|\varphi\rangle$ at $t=0$,
the time evolution can be found by writing $|\varphi\rangle$ in terms of 
the energy eigenstates. One gets the probability of finding the system in
$|\varphi\rangle$ at time $t$ as,
\begin{align}
P_\varphi(t)&=|\langle\varphi|\psi(t)\rangle|^2=\left|\int|\alpha(E')|^2e^{-iE't}dE'\right|^2\,,
\end{align}
which can be evaluated exactly for the Wigner-Weisskopf case to get the
exponential decay behavior predicted by the Fermi golden rule:
\begin{align}
    P_\varphi(t)=e^{-2\pi W^2 t}\,.
\end{align}
A similar formula can be derived for the resonant coupling case, which we
defer to Section \ref{6}.

\section{Generalizing the Bixon-Jortner system}\label{3}
\subsection{The general formalism}\label{3.1}
We will start with a general system describing a coupling between a discrete 
state and a quasi-continuum, and will then specialize to the particular 
case we will investigate. Once again, we use the notation of 
\cite{tannoudji1992atom}. 
Our total Hamiltonian is $H = H_{0} + V$, where $H_0$ denotes the unperturbed piece whose eigenvectors consist of an infinite sequence
of states $|k\rangle$ for $k=0, \pm 1, \pm 2, \ldots$,
and a separate, discrete state $|\varphi\rangle$, and $V$ denotes the coupling between the 
discrete state and the infinite sequence. As in the Bixon-Jortner system, 
we will say that the states $|k\rangle$ form a
quasi-continuum or ``ladder''.
In this basis, $H_0$ is given by,
\begin{eqnarray*}
H_0 |k\rangle &=& E_k |k\rangle \\
H_0 |\varphi \rangle &=& E_{\varphi}|\varphi\rangle \,.
\end{eqnarray*}
The perturbation $V$ couples the ladder to the discrete state:
\begin{eqnarray*}
\langle k|V| \varphi \rangle &=& v_k = \langle \varphi|V| k \rangle^*\\
\langle k|V| k'\rangle &=& 0\\
\langle \varphi |V| \varphi\rangle &=&0\,.
\end{eqnarray*}
Below, we will assume that $E_k = k\delta$ where $\delta$ denotes the uniform separation 
between the ladder states.

We will denote the eigenvalues and the eigenvectors
of the full Hamiltonian $H=H_0 + V$ by $E_{\mu}$ and
$|\psi_{\mu}\rangle$, respectively:
\begin{equation}\label{general-eig-eq}
H\ket{\psi_\mu} = E_\mu\ket{\psi_\mu}\,.
\end{equation}
To seek these eigenvalues/eigenvectors in terms of those of $H_0$, we project 
Equation \eqref{general-eig-eq} to $|\varphi\rangle$ and $|k\rangle$, respectively, and use the resolution of identity,
\begin{equation*}
I = |\varphi\rangle\langle\varphi| + \sum_{k} |k\rangle\langle k|\,.
\end{equation*}
This gives,
\begin{eqnarray}
E_{\varphi} \langle \varphi|\psi_{\mu}\rangle + \sum_k v_k^* \bra{k}\ket{\psi_\mu} &=& E_\mu\bra{\varphi}\ket{\psi_\mu}\label{phi-eq}\\
E_k \bra{k}\ket{\psi_\mu} + v_k\bra{\varphi}\ket{\psi_\mu}&=& E_\mu\bra{k}\ket{\psi_\mu}\,,\label{k-eq}
\end{eqnarray}
respectively, for the two projections.
Equation \eqref{k-eq} gives
\begin{equation}\label{k-phi-inner-rel}
\bra{k}\ket{\psi_\mu} = \bra{\varphi}\ket{\psi_\mu} \frac{v_k}{E_\mu - E_k}\,,
\end{equation}
and substituting in \eqref{phi-eq}, one gets
\begin{equation}\label{eigval}
E_\mu = E_{\varphi} + \sum_k \frac{|v_k|^2}{E_\mu - E_k} 
\end{equation}
For given $k$-dependent $E_k$ and $v_k$, and a given value for $E_{\varphi}$,
solving this equation would give the eigenvalues $E_{\mu}$ of the total Hamiltonian 
$H$. We call Equation \eqref{eigval} the eigenvalue equation.
Using \eqref{k-phi-inner-rel} and the normalization
condition
\begin{equation*}
\sum_k  \abs{\bra{k}\ket{\psi_\mu}}^2 + \abs{\bra{\varphi}\ket{\psi_\mu}}^2 = 1
\end{equation*}
and an overall choice of phase, we get the eigenvectors $|\psi_{\mu}\rangle$ 
in terms of the eigenvalues $E_{\mu}$,
\begin{eqnarray}
\bra{\varphi}\ket{\psi_\mu} &=&
\frac{1}{\bigg[1+\sum_{k'}{\frac{|v_{k'}|^2}{(E_\mu - E_{k'})^2}}\bigg]^\frac{1}{2}}\label{eigvec-phi}\\
\bra{k}\ket{\psi_\mu} &=& 
\frac{v_k/{(E_\mu - E_k)}}{\bigg[1+\sum_{k'}{\frac{|v_{k'}|^2}{(E_\mu - E_{k'})^2}}\bigg]^\frac{1}{2}}\label{eigvec-k}
\end{eqnarray}

The equations \eqref{eigval}, \eqref{eigvec-phi}, and \eqref{eigvec-k} give us
the eigenvalues and the eigenvectors of the coupled Hamiltonian in terms of the 
eigenvalues and eigenvectors of the uncoupled Hamiltonian and the coupling constants. 
Of course, obtaining analytic solutions for these is another matter.

\subsection{The Lorentzian Bixon-Jortner model}\label{3.2}

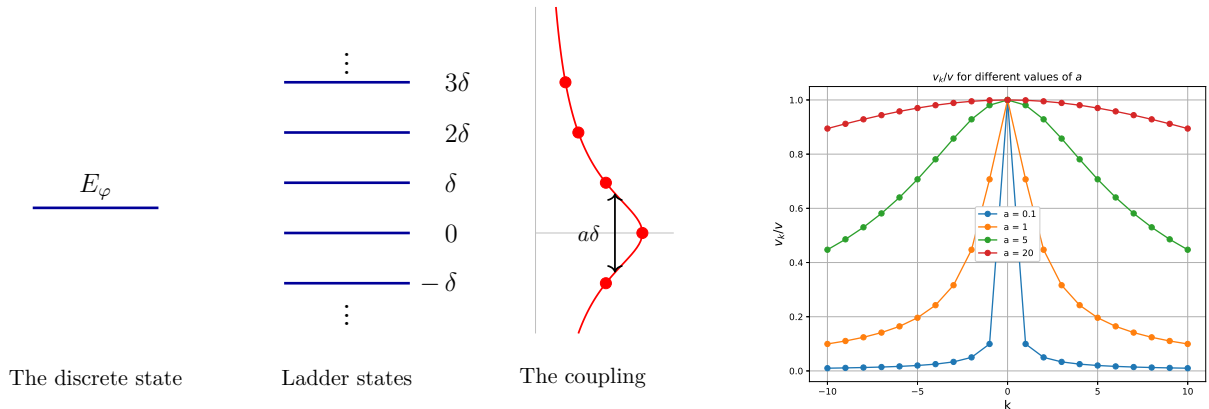
\begin{figure}[h!]\label{fig:coupling-plots}
    \centering
    \begin{subfigure}[b]{0.55\textwidth}
        \centering
        \resizebox{\textwidth}{!}{%
        \begin{tikzpicture}[
            level/.style={draw=blue!60!black, very thick},
            desc_label/.style={text width=3.5cm, align=center},
            elabel/.style={anchor=west, font=\large}
        ]
            \def\spacing{0.8} 
            \def\curvA{0.7}   
            \def\curvM{1.7}   

            \foreach \y in {0, 1, 2, 3, 4} {
                \pgfmathtruncatemacro{\n}{\y-1} 
                \draw[level] (1, \y*\spacing) -- (3, \y*\spacing); 
                \node[elabel] at (3.4, \y*\spacing) { 
                    \ifnum\n=0 $0$\fi
                    \ifnum\n=1 $\delta$\fi
                    \ifnum\n>1 $\n\delta$\fi
                    \ifnum\n=-1 \makebox[0pt][r]{$-$}$\delta$\fi
                    \ifnum\n<-1 \pgfmathtruncatemacro{\nabs}{abs(\n)}\makebox[0pt][r]{$-$}$\nabs\delta$\fi
                };
            }
            \node[font=\Large] at (2, 4.5*\spacing) {$\vdots$}; 
            \node[font=\Large] at (2, -0.5*\spacing) {$\vdots$}; 
            \node[desc_label] at (2, -1.5) {Ladder states}; 

            \draw[level] (-3, 1.5*\spacing) -- (-1, 1.5*\spacing) node[midway, above, font=\large] {$E_{\varphi}$};
            \node[desc_label] at (-2, -1.5) {The discrete state};

            \draw[gray!50] (5, 1*\spacing) -- (5 + \curvM + 0.5, 1*\spacing);
            \draw[gray!50] (5, -1*\spacing) -- (5, 5.5*\spacing);

            \draw[thick, red, domain=-1:5.5, samples=100, variable=\t]
                plot ({5 + \curvM/sqrt(1 + ((\t*\spacing - 1*\spacing)^2)/(\curvA^2))}, {\t*\spacing});

            \foreach \y in {0, 1, 2, 3, 4} {
                \pgfmathsetmacro{\dotX}{5 + \curvM/sqrt(1 + ((\y*\spacing - 1*\spacing)^2)/(\curvA^2))}
                \pgfmathsetmacro{\dotY}{\y*\spacing}
                \filldraw[red] (\dotX, \dotY) circle (2.5pt);
            }

            \pgfmathsetmacro{\arrowL}{1.8 * \curvA}
            \pgfmathsetmacro{\arrowX}{5 + \curvM / sqrt(1 + ((\arrowL/2)^2) / (\curvA^2))}
            \draw[<->, thick] (\arrowX, {1*\spacing - \arrowL/2}) -- (\arrowX, {1*\spacing + \arrowL/2}) node[midway, left, xshift=-2pt] {$a\delta$};

            \node at (5.75, -1.5) {The coupling};

        \end{tikzpicture}
        }
        \caption{Schematic representation of the energy levels and the coupling. The ladder
        states have unperturbed energy $E_n = n\delta$, and the discrete state has energy $E_{\varphi}$. 
        The coupling between the state $|\varphi\rangle$ and $|n\rangle$ is a curve with a maximum
        at $n=0$, and width $a\delta$, where $a$ is the dimensionless width of the coupling curve.}
        \label{fig:energy_schematic}
    \end{subfigure}
    \hfill 
    \begin{subfigure}[b]{0.4\textwidth}
        \centering\label{fig:generalized-coupling-plots}
        \includegraphics[width=\textwidth]{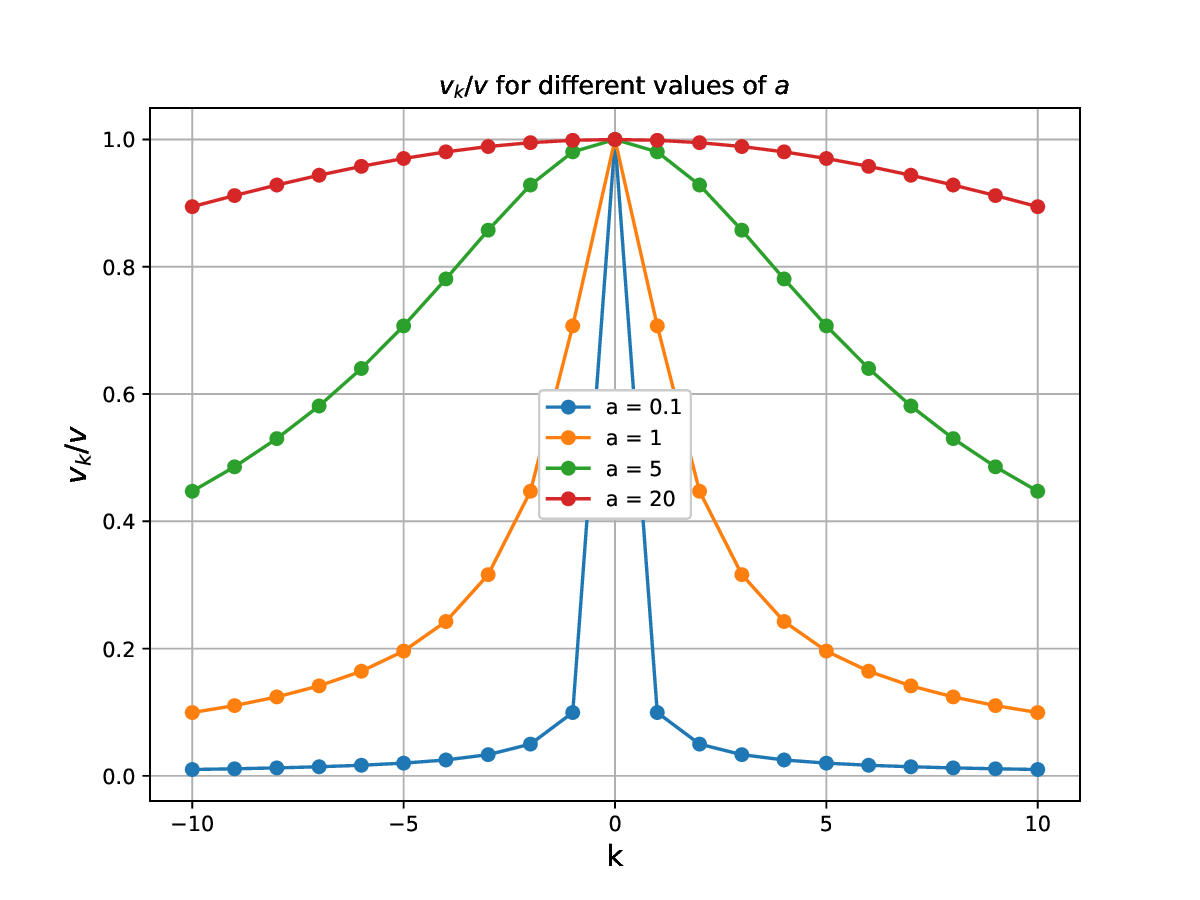}
        \caption{Plot of the coupling strength $v_k$ with respect to $k$ for a few values of $a$.
        For large $a$, we get near-constant coupling, for small $a$, we get a coupling that is 
        nonzero only for $k=0$.}
        \label{fig:coupling_plot}
    \end{subfigure}
    \caption{The Lorentzian generalization of the Bixon-Jortner system.}
    \label{fig:combined_figures}
\end{figure}

In the Bixon-Jortner (BJ) model of Section \ref{2.1}, 
one has $E_k = k\delta$ and a constant value of the coupling, $v_k = v$. Here, 
we generalize this to a Lorentzian profile 
with a given resonance width. More explicitly, we make
the following choice for the couplings and the
unperturbed eigenvalues of the quasi-continuum:
\begin{equation}\label{eq:system-defn}
\begin{aligned}
E_{k} &= k\delta \\ 
v_{k} &= v\frac{\gamma}{\sqrt{\gamma^2 + E_k^2}} = \frac{v}{\sqrt{1 + (\frac{k}{a})^2}}\,,
\end{aligned}
\end{equation}
where $v$ is a real number,\footnote{As in the Bixon-Jortner case,
a complex generalization is trivial and doesn't bring anything new---the
phase of $v$ comes up only in Equation \eqref{eigvec-k}.} and $a = \gamma/\delta$ is a dimensionless variable measuring the width of the coupling  $v_k$ in units of $\delta$.
See Figure \ref{fig:combined_figures} for a visual representation.
The quasi-continuum still forms a 
uniform ladder as in the BJ system, but the coupling now has a peak at $k=0$ with dimensionless width $a$.


In the limit $a\to 0$, the only ladder state coupled to $|\varphi\rangle$ will be $|k=0\rangle$,
all the others getting decoupled. Thus, this limit describes a two-state (Rabi) system
with the basis $|\varphi\rangle$ and $|k=0\rangle$,
together with a decoupled ladder of states $|k\rangle$, $k=\pm 1, \pm 2, \ldots$. 
In the opposite limit of $a\to\infty$, all the couplings get the same value $v_k = v$. This is 
the BJ limit of our system. Thus, the system \eqref{eq:system-defn} 
interpolates between the BJ system and the Rabi two-state system (with a decoupled ladder) as the parameter $a$ is varied between $\infty$ and $0$.
Some examples plots of $v_k$
can be seen in Figure~\ref{fig:combined_figures}.

Another adjustable parameter is the spacing, $\delta$ 
(or more appropriately,
the dimensionless measure of it $\delta/v$). The limit $\delta \to 0$ corresponds to a continuum of energy levels. In the generic case, this gives the Lorentzian 
continuum reviewed in Section \ref{2.3}, and in the case of large $a$, it gives the 
Weisskopf-Wigner model described in the same section. In short, the model 
\eqref{eq:system-defn} unifies the Rabi, BJ, Weisskopf-Wigner, and the Lorentzian version of the Fano system, and reduces to these models in special limits.

The time evolution of the Rabi system is known to be oscillatory, whereas
when the BJ system is started in the discrete state, it has decays and 
revivals. The Weisskopf-Wigner system has an exponential decay behavior
with the decay rate given exactly by the Fermi golden rule, whereas the 
Lorentzian Fano system has oscillatory decays. As we will demonstrate in Section \ref{6} 
the Lorentzian BJ model is rich enough to cover all these behaviors.

\section{The solution}\label{4}
\subsection{Eigenvalues}

\begin{figure}[htp]\label{eigenval}
    \centering
    \includegraphics[width=0.55\textwidth]{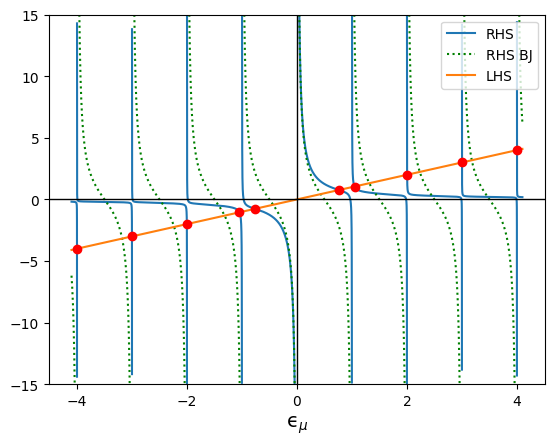}
    \caption{The left hand side (LHS) and the right hand side (RHS) of the
    transcendental eigenvalue equations \eqref{final-eig} and \eqref{BJ_eigval}. 
    }
    \label{fig:a_zero_limit_eigval}
\end{figure}

Plugging \eqref{eq:system-defn} into \eqref{eigval}, we get 
\begin{equation}    
 E_\mu = E_\varphi + \sum_k \frac{v^2}{1 + \frac{k^2}{a^2}} \frac{1}{E_\mu - k \delta} 
\end{equation}
Defining the dimensionless forms of the energy eigenvalues $\epsilon_{\mu} = E_{\mu}/\delta$,
$\epsilon_{\varphi} = E_{\varphi}/\delta$, we get
\begin{equation}\label{eig-transcen}
 \epsilon_{\mu} = \epsilon_{\varphi} + 
 \frac{v^2}{\delta^2}\sum_k \frac{1}{1 + \frac{k^2}{a^2}} \frac{1}{\epsilon_{\mu} - k} 
\end{equation}
This transcendental equation specifies the energy eigenvalues $\epsilon_{\mu}$ of the coupled system.
To get a closed form version of \eqref{eig-transcen}, we need to evaluate the
infinite sum
\begin{equation}\label{eq:s1-defn}
S_1(\epsilon_{\mu}, a) = \sum_{k=-\infty}^{\infty} \frac{1}{1 + \frac{k^2}{a^2}} \frac{1}{\epsilon_{\mu} - k} 
\end{equation}
We accomplish this by using the Mittag-Leffler theorem. 
Expanding into partial fractions,
\begin{equation}\label{eq:partial}
S_1(\epsilon_{\mu}, a) =  
\frac{a^2}{a^2 + \epsilon_{\mu}^2}\sum_{k=-\infty}^{\infty} \left(\frac{ \epsilon_{\mu}}{a^2 + k^2} + \frac{k}{a^2 + k^2} +   \frac{1}{\epsilon_{\mu} - k}\right)
\end{equation}
Now, $S_1(\epsilon_{\mu}, a)$ as defined in \eqref{eq:s1-defn} is absolutely convergent,
and we are free to evaluate it using the ``principal value'' approach of $\lim_{N\to\infty}\sum_{k=-N}^{N}$
This makes it possible to evaluate \eqref{eq:partial} term by term:
\begin{equation}\label{eq:principal-val-sum}
S_1(\epsilon_{\mu}, a) =  
\frac{a^2}{a^2 + \epsilon_{\mu}^2}\left[\lim_{N\to\infty}\epsilon_{\mu}\sum_{k=-N}^{N} \frac{ 1}{a^2 + k^2} + \lim_{N\to\infty}\sum_{k=-N}^{N} \frac{k}{a^2 + k^2} + \lim_{N\to\infty}\sum_{k=-N}^{N} \frac{1}{\epsilon_{\mu} - k}\right]
\end{equation}
The second sum vanishes. The first sum is also absolutely convergent, and is given by 
\begin{equation}
\sum_{k=-\infty}^{\infty} \frac{1}{a^2 + k^2} = \frac{\pi}{a} \coth(\pi a) 
\end{equation}
The third sum gives,
\begin{equation}
\lim_{N\to\infty}\sum_{k=-N}^{N} \frac{1}{\epsilon_{\mu}- k} = \pi \cot(\pi \epsilon_{\mu})\,,
\end{equation}
see, e.g., \cite{cartan1995elementary}, page 153.
Note that this sum also comes up in the Bixon-Jortner model, but it only becomes convergent
using the symmetric limit, which means that the infinite BJ system exists only as the limit of a \emph{symmetric}
finite ladder. This point is glossed over in \cite{tannoudji1992atom2}. 
Combining the terms, we get,
\begin{equation}\label{s1-eq}
S_1(\epsilon_{\mu}, a) = \pi\left[
\frac{ \cot(\pi \epsilon_{\mu}) +  \epsilon_{\mu} \coth(\pi a)/a}{1 + (\epsilon_{\mu}/a)^2} \right]\,.
\end{equation}
Substituting this in \eqref{eig-transcen}, 
we see that the dimensionless 
energy eigenvalues $\epsilon_{\mu}$  are given by the 
solutions to the transcendental equation,
\begin{equation}\label{final-eig}
\epsilon_{\mu} = \epsilon_{\varphi} + \frac{\pi v^2}{\delta^2} \left[
\frac{ \cot(\pi \epsilon_{\mu}) +  \alpha(a) \epsilon_{\mu} }{1 + 
(\epsilon_{\mu}/a)^2} \right]\,,
\end{equation}
where $\alpha(a) = \coth(\pi a)/a$ is a term independent of energy.
In Figure \ref{fig:a_zero_limit_eigval}, we show the left hand side (LHS) and the right hand side (RHS)
of equation \eqref{final-eig}, together with the RHS of the Bixon-Jortner
eigenvalue equation \eqref{BJ_eigval}. See Section \ref{visncomm} for further
visualization of the eigenvalue equation as the parameters are varied.

\subsection{Eigenvectors}

We next use \eqref{eigvec-phi} and \eqref{eigvec-k} to obtain the eigenvectors.
Once again, working with the dimensionless $\epsilon_{\mu} = E_{\mu}/\delta$, we
have,
\begin{eqnarray}\label{eq:eigvec-varphi-component}
    \langle\varphi|\psi_{\mu}\rangle &=& \frac{1}{\left[1 + \frac{v^2}{\delta^2}
S_2(\epsilon_{\mu}, a) \right]^{1/2}}\\
\langle k|\psi_{\mu}\rangle &=& \langle\varphi|\psi_{\mu}\rangle \frac{v_k}{\delta(\epsilon_{\mu}-k)}
\label{eq:eigvec-k-component}
\end{eqnarray}
where
\begin{equation}
    S_2(\epsilon_{\mu}, a) = \sum_{k=-\infty}^{\infty} \frac{1}{1 + (k/a)^2}\frac{1}{(\epsilon_{\mu} - k)^2}\,.
\end{equation}
To evaluate this sum, simply note that 
$S_2(\epsilon_{\mu}, a) =
-\frac{\partial S_1(\epsilon_{\mu}, a)}{\partial \epsilon_{\mu}}$ where $S_1$ is given in \eqref{s1-eq}.
Thus,
\begin{align}
    S_2(\epsilon_{\mu}, a)
    &= \frac{\pi}{1 + (\epsilon_{\mu}/a)^2}\left[\pi\csc^2(\pi\epsilon_{\mu}) - \alpha(a)
+ \frac{2\epsilon_{\mu}}{a^2}\frac{\cot(\pi \epsilon_{\mu}) + \alpha(a)\epsilon_{\mu} }{1 + (\epsilon_{\mu}/a)^2}
\right]\,,\label{eq:s2-trig-version}
\end{align}
where $\alpha(a) = \coth{(\pi a)}/a$ as before.

It is also useful to obtain the sum $S_2(\epsilon_{\mu}, a)$ and hence
the eigenvectors as purely rational functions of $\epsilon_{\mu}$.  
This can be accomplished by obtaining the cotangent term in 
\eqref{eq:s2-trig-version} as a rational function
from the eigenvalue equation \eqref{final-eig}.
We denote $\cot{\pi \epsilon_{\mu}}$ obtained this way 
by $\mathcal{C} = \mathcal{C}(\epsilon_{\mu}, a)$,
which is given by,
\begin{equation}\label{eq:c-equation}
 \cot{\pi\epsilon_{\mu}} = \mathcal{C}(\epsilon_{\mu}, a)= \frac{\delta^2}{\pi v^2}\left(1 + (\epsilon_{\mu}/a)^2\right)(\epsilon_{\mu} - \epsilon_{\varphi}) -\epsilon_{\mu} \alpha(a)
\end{equation}
Substituting in \eqref{eq:s2-trig-version}, we get 
\begin{equation}\label{eq:s2-rational-version}
    S_2(\epsilon_{\mu}, a) = \frac{\pi}{1 + (\epsilon_{\mu}/a)^2}\left[
    \pi(1 + \mathcal{C}^2) - \alpha(a) + \frac{2\epsilon_{\mu}}{a^2}
    \frac{\epsilon_{\mu} \alpha(a) + \mathcal{C}}{1 + (\epsilon_{\mu}/a)^2}
    \right]\,.
\end{equation}
Note that this is now a rational function of $\epsilon_{\mu}$, which (as opposed to \eqref{eq:s2-trig-version}), also has 
an explicit $v$-dependence through $\mathcal{C}$.

\section{The behavior of the solution}\label{5}
We next investigate the behavior of the eigenvalues and eigenvectors in various limits. 
For the original Hamiltonian given by \eqref{eq:system-defn},
the limits we consider are,
\begin{enumerate}
    \item The $a\to\infty$ limit, which makes the system approach the BJ system,
\item The $a\to 0$ limit which makes the system approach a two-state (Rabi) 
system (consisting of $|0\rangle$ and $|\varphi\rangle$) 
and a decoupled subsystem (consisting of the remaining $|n\ne 0\rangle$),
\item The $\delta/v \to 0$ limit which makes the system approach a true continuum (this limit is taken while keeping $\Gamma = 2\pi v^2/\delta$ and $\gamma = a\delta$ constant). 
\end{enumerate}
Below, we will make use of the fact that the function $\alpha(a)$ defined above vanishes in both the small $a$ and the 
large $a$ limit, with the limiting behavior,
\begin{align}
\alpha(a) =\frac{\coth \pi a}{a} \approx \begin{cases}1/a & a\to\infty\\ 1/(\pi a^2)& a\to0\end{cases}\,.
\end{align}

\subsection{The $a\to\infty$ (BJ) limit}\label{BJ-limit}

\subsubsection{Eigenvalues}
Using $a\to\infty$ in \eqref{final-eig}, we get 
\begin{equation}\label{BJ-limit}
\epsilon_{\mu} \approx \epsilon_{\varphi} + \frac{\pi v^2}{\delta^2} 
\cot(\pi \epsilon_{\mu}) 
\end{equation}
As expected, this is precisely the eigenvalue equation \eqref{BJ_eigval} for the Bixon-Jortner system.

\subsubsection{Eigenvectors}



For $a\to\infty$, using $\alpha(a)\approx 1/a$ on \eqref{eq:c-equation} and \eqref{eq:s2-rational-version},
we get 
\begin{align}
    \mathcal{C}(\epsilon_\mu, a)&\approx\frac{\delta^2}{\pi v^2}(\epsilon_\mu-\epsilon_\varphi)\\
    S_2(\epsilon_\mu, a)&\approx\pi^2\left(1+\frac{\delta^4}{\pi^2v^4}(\epsilon_\mu-\epsilon_\varphi)^2\right)\,.
\end{align}
Using \eqref{eq:eigvec-varphi-component} and \eqref{eq:eigvec-k-component},
these give the eigenvectors in this limit as
\begin{align}
    \langle\varphi|\psi_\mu\rangle&=\frac{1}{\left[1+\frac{v^2\pi^2}{\delta^2}+\frac{\delta^2}{v^2}(\epsilon_\mu-\epsilon_\varphi)^2\right]^{1/2}}\\
    \langle k|\psi_\mu\rangle&=\frac{v}{\delta(\epsilon_\mu-k)} \langle\varphi|\psi_\mu\rangle\,.
\end{align}
These are exactly the Bixon-Jortner eigenvector formulas
\eqref{bj-eigvecs-1}-\eqref{bj-eigvecs-2}
(see also Equation (17)
on page 55 of \cite{tannoudji1992atom}, noting that they use $E_{\varphi}=0$).



\subsection{The $a\to 0$ (decoupled Rabi) limit}\label{decoupled-rabi}

\subsubsection{Eigenvalues}
In the limit $a\to 0$, there are two cases. If $\epsilon_{\mu}$ is not close to an integer, 
then the cotangent term in the eigenvalue equation \eqref{final-eig} can be ignored and the equation becomes 
\begin{equation}
\epsilon_{\mu} \approx \epsilon_{\varphi} + \frac{v^2}{\delta^2}\frac{1}{\epsilon_{\mu}}
\end{equation}
which gives the approximate eigenvalues as
\begin{equation}\label{eq:rabi-limit-eigvals}
    \epsilon_{\mu\pm} = \frac{1}{2}\Big(\epsilon_{\varphi} \pm \sqrt{\epsilon_{\varphi}^2 + 4v^2/\delta^2}\Big) \,.
\end{equation}
These are simply the energy eigenvalues of a two-state (Rabi) system with Hamiltonian,
\begin{align}
    H_{\varphi,0} = \begin{bmatrix}
        \epsilon_{\varphi} & v/\delta \\
        v/\delta & 0
    \end{bmatrix}\,,
\end{align}
acting on the subspace spanned by the original discrete state $|\varphi\rangle$
and the single ladder state $|k=0\rangle$; see \eqref{eq:rabi-h-e2-zero}. 
Once again, this is the expected behavior
considering the form of the coupling in \eqref{eq:system-defn}---in the limit $a\to 0$,
only the $k=0$ state is coupled to $|\varphi\rangle$ .

The remaining set of solutions to the eigenvalue equation \eqref{final-eig} in the limit $a\to 0$
consists of (approximate) nonzero
integers. To see how the solutions approach integers in this limit, let
$\epsilon_{\mu} = n + \Delta$, with $\Delta$ small. Then, for $a\to 0$, \eqref{final-eig} becomes, approximately,
\begin{equation}
    n \approx \epsilon_{\varphi} + \frac{v^2}{\delta^2} \Big(
    \frac{a^2}{n^2 \Delta} + \frac{1}{n}\Big)
\end{equation}
 whose solution for $\Delta$ is
 \begin{equation}\label{Delta-propa}
     \Delta = \frac{v^2 a^2}{\delta^2 n^2}\left[\frac{1}{
     n - \epsilon_{\varphi} - v^2/(n \delta^2)}\right]
 \end{equation}
which is $\mathcal{O}(a^2)$ as $a\to 0$. This gives the convergence rate 
of $\epsilon_{\mu}$ to integer values as $a\to 0$.


In short, in the $a\to 0$ limit, the system decouples into two pieces: a two-state Rabi system
spanned by $|\varphi\rangle$ and $|k=0\rangle$, and a decoupled infinite ladder 
$|k\rangle$, $k=\pm 1, \pm 2, \ldots$ with a missing step.


\subsubsection{Eigenvectors}

Once again, there are two cases. If $\epsilon_\mu$ is not close to an integer,
its approximate values are given by \eqref{eq:rabi-limit-eigvals}, and $\cot(\pi \epsilon_{\mu\pm})$ and $\csc(\pi 
\epsilon_{\mu\pm})$ have the corresponding limiting values. 
Using $\alpha(a)\approx 1/(\pi a^2)$ as ${a\to 0}$ and working to lowest order in $a$, we get from 
\eqref{eq:s2-trig-version},
\begin{align}
    S_2(\epsilon_\mu,a)\approx\frac{1}{\epsilon^2_\mu}\,,
\end{align}
which gives the $|\varphi\rangle$ component of the eigenvectors via \eqref{eq:eigvec-varphi-component} 
as
\begin{align}
    \langle \varphi|\psi_\mu\rangle=\frac{1}{\left[1+\frac{v^2}{\delta^2}\frac{1}{\epsilon^2_\mu}\right]^{1/2}}\,.\label{eq:a-zero-eigvec-phi}
\end{align}
For the $\langle k|\psi_{\mu}\rangle$ components, we have from \eqref{eq:eigvec-k-component},
\begin{align}
    \langle k|\psi_\mu\rangle&=\frac{v}{\delta(\epsilon_\mu-k)\sqrt{1 + \frac{k^2}{a^2}}}\langle\varphi|\psi_{\mu}\rangle\,,
\end{align}
which gives, as $a\to 0$, $\langle k|\psi_\mu\rangle \approx 0$ for $k\ne 0$ and
\begin{align}
    \langle 0|\psi_\mu\rangle&=\frac{v}{\delta\epsilon_\mu}\frac{1}{\left[1+\frac{v^2}{\delta^2}\frac{1}{\epsilon^2_\mu}\right]^{1/2}}\,. \label{eq:a-zero-eigvec-zero}
\end{align}
Thus, we get the nonzero components of $|\psi_{\mu}\rangle$ as,
\begin{align}\label{eq:rabi-eigvecs-simple-1}
    \langle 0|\psi_\mu\rangle=\frac{v/\delta}{\left[\epsilon_{\mu}^2+(v/\delta)^2\right]^{1/2}}\,, \quad{}
    \langle \varphi|\psi_\mu\rangle=\frac{\epsilon_{\mu}}{\left[\epsilon_{\mu}^2+(v/\delta)^2\right]^{1/2}}\,,
\end{align}
for the case where $\epsilon_{\mu}$ is not close to an integer.
Using \eqref{eq:rabi-limit-eigvals} for $\epsilon_{\mu} = E_{\mu}/\delta$, equations \eqref{eq:rabi-eigvecs-simple-1} are precisely the eigenvectors of a two-state Rabi
system with a Hamiltonian
\begin{align}
    H_{\text{Rabi}} = \begin{bmatrix}0 & v/\delta \\ v/\delta & \epsilon_{\varphi}
    \end{bmatrix}
\end{align}
see \eqref{eq:rabi-eigvecs-ours-real-v}. 
Incidentally, the formulas in \eqref{eq:rabi-eigvecs-simple-1} 
give a simple representation of the Rabi eigenvectors in terms of the eigenvalues
which we hadn't encountered in the literature.

If $\epsilon_\mu$ is close to an integer, $\epsilon_\mu=n+\Delta$, according to \eqref{Delta-propa},  $\Delta$  is 
approximately proportional to $a^2$, $\Delta\approx \beta a^2$. Thus we can write $\cot(\pi \epsilon_\mu)
=\cot(\pi n+\pi\Delta)= \cot(\pi \Delta)\approx\frac{1}{\pi \Delta}$, and similarly for $\csc^2(\pi \epsilon_{\mu})$,
and get,
\begin{align}
    \cot(\pi\epsilon_{\mu})^2\approx\frac{1}{\pi \beta a^2}\,,\quad{}
    \csc^2(\pi \epsilon_{\mu})\approx \frac{1}{\pi^2\beta^2a^4}\,.
\end{align}
These give $S_2(\epsilon_\mu,a)$ as,
\begin{align}
    S_w(\epsilon_\mu,a)&\approx\frac{\pi}{n^2}\frac{1}{\frac{1}{a^2}}\left[\frac{1}{\pi \beta^2 a^4}-\frac{1}{\pi a^2} \frac{2n}{a^2}\frac{\frac{1}{\pi\beta a^2}+\frac{n}{\pi a^2}}{\frac{n^2}{a^2}}   \right]\\
    &\approx\frac{1}{n^2\beta^2 a^2}\,.
\end{align}
Finally, we get the components of the eigenvector $|\psi_{\mu}\rangle$ corresponding to a 
near-integer eigenvalue $\epsilon_{\mu} = n + \Delta$ with $\Delta \approx \beta a^2$ as,
\begin{align}
    \langle\varphi|\psi_\mu\rangle\approx\frac{1}{\left[1+\frac{v^2}{\delta^2}\frac{1}{n^2\beta^2a^2}\right]^{1/2}}&\approx 0\\
        \langle k|\psi_\mu\rangle \approx \frac{v}{\frac{\delta k}{a}(n-k)}\frac{1}{\frac{v}{\delta}\frac{1}{n\beta a}}&\approx \begin{cases}
            0 & \text{if } k\ne n\\
            1 & \text{if } k=n
        \end{cases}\,.
\end{align}
In short, in the limit $a\to 0$, the eigenvectors corresponding to eigenvalues that are near integers
are given approximately by the unperturbed eigenvectors $|k\rangle$ with $k=\pm 1, \pm 2, \ldots$, 
and the remaining eigenvectors are given by a linear combination of $|\varphi\rangle$ and $|0\rangle$
the coefficients being given in terms of the appropriate Rabi eigenvectors via \eqref{eq:rabi-eigvecs-ours-real-v}. 
We emphasize that while the eigenvalues corresponding to 
$n\ne 0$ are close to integers $a\to 0$ and are thus independent of $v$,
the eigenvalues in the space spanned by $|n=0\rangle$
and $|\varphi\rangle$ do depend on $v$ in this limit.

\subsection{The $\delta \to 0$ (continuum) limit}

In the continuum limit where the spacing $\delta\to 0$, both 
the original, unperturbed energy eigenvalues $E_k = \delta k$ and the  eigenvalues $E_{\mu}$ of the full Hamiltonian become 
continuous. In order to have a well-defined limiting behavior 
in this case,  we keep the two quantities $\Gamma = \frac{2\pi v^2}{\delta}$ and $\gamma = a\delta$ fixed as $\delta\to 0$. 
This amounts to keeping the decay rate given by the Fermi golden rule approximation and the width of the resonance coupling constant as $\delta\to 0$. 
Instead of the dimensionless energy values $\epsilon_{\mu}$ and $\epsilon_{\phi}$, in this section
we work with their dimensionful versions $E_{\mu} = \epsilon_{\mu}\delta$ and $E_{\varphi} = \epsilon_{\mu}\delta$.

We focus on the component $\langle\varphi|\psi_{\mu}\rangle$ of the eigenvector $|\psi_{\mu}\rangle$. Via \eqref{eq:c-equation}, 
we get to lowest order in $\delta$,
\begin{align}
    \mathcal{C} \approx \frac{2}{\Gamma}\left(1+\left(\frac{E_\mu}{\gamma}\right)^2\right)(E_\mu-E_\varphi)-\frac{E_\mu}{\gamma} \,.
\end{align}
Substituting this in \eqref{eq:s2-rational-version} 
to get $S_2$ and working once again to lowest order in $\delta$, we get,
\begin{align}
    S_2 \approx \frac{\pi^2}{1+\left(\frac{E_\mu}{\gamma}\right)^2}\left[1+\left(\frac{2}{ \Gamma}\left(1+\left(\frac{E_\mu}{\gamma}\right)^2\right)(E_\mu-E_\varphi)-\frac{E_\mu}{\gamma}\right)^2\right]\,.
\end{align}
Finally, substituting in \eqref{eq:eigvec-varphi-component}  
and defining $W^2 = \Gamma \gamma /2$
we get after some manipulation,
\begin{align}\label{prob-density}
    |\langle \varphi|\psi_\mu\rangle|^2 = \frac{W^2\gamma}{\pi} \frac{\delta}{(E_\mu(E_\mu-E_\varphi)-W^2)^2+\gamma^2(E_\mu-E_\varphi)^2}\,.
\end{align}
This matches \eqref{alphaabs} in Section \ref{2.3} and formula 6.5.45 in \cite{barnett2002methods65}. 
Note that the remaining $\delta$ in  can either be absorbed in a redefinition (``renormalization'') of the eigenstates $|\psi_{\mu}\rangle$
in the continuum limit to get a $\delta$-function 
normalization, or can be explicitly dealt with when summing 
over $\mu$, using $\sum_{\mu}f(E_{\mu}) \delta \to \int f(E) 
dE$. We will follow this latter approach below in our 
discussion of the decay of the discrete state in this limit.

\subsection{Visualization and commentary}\label{visncomm}

In Figure \ref{eigval-graphs}, 
we present a graphical representation of the transcendental 
eigenvalue equation \eqref{final-eig} by plotting the two sides of this equation for various parameter settings.
Each intersection of these two sides corresponds to an eigenvalue.
The figure demonstrates the limiting behaviors we discussed above,
where the Rabi, Bixon-Jortner, and decoupled ladder pieces become relevant
in specified limits.






\begin{figure}[htp]\label{eigval-graphs}
    \centering
    \includegraphics[width=0.9\textwidth]{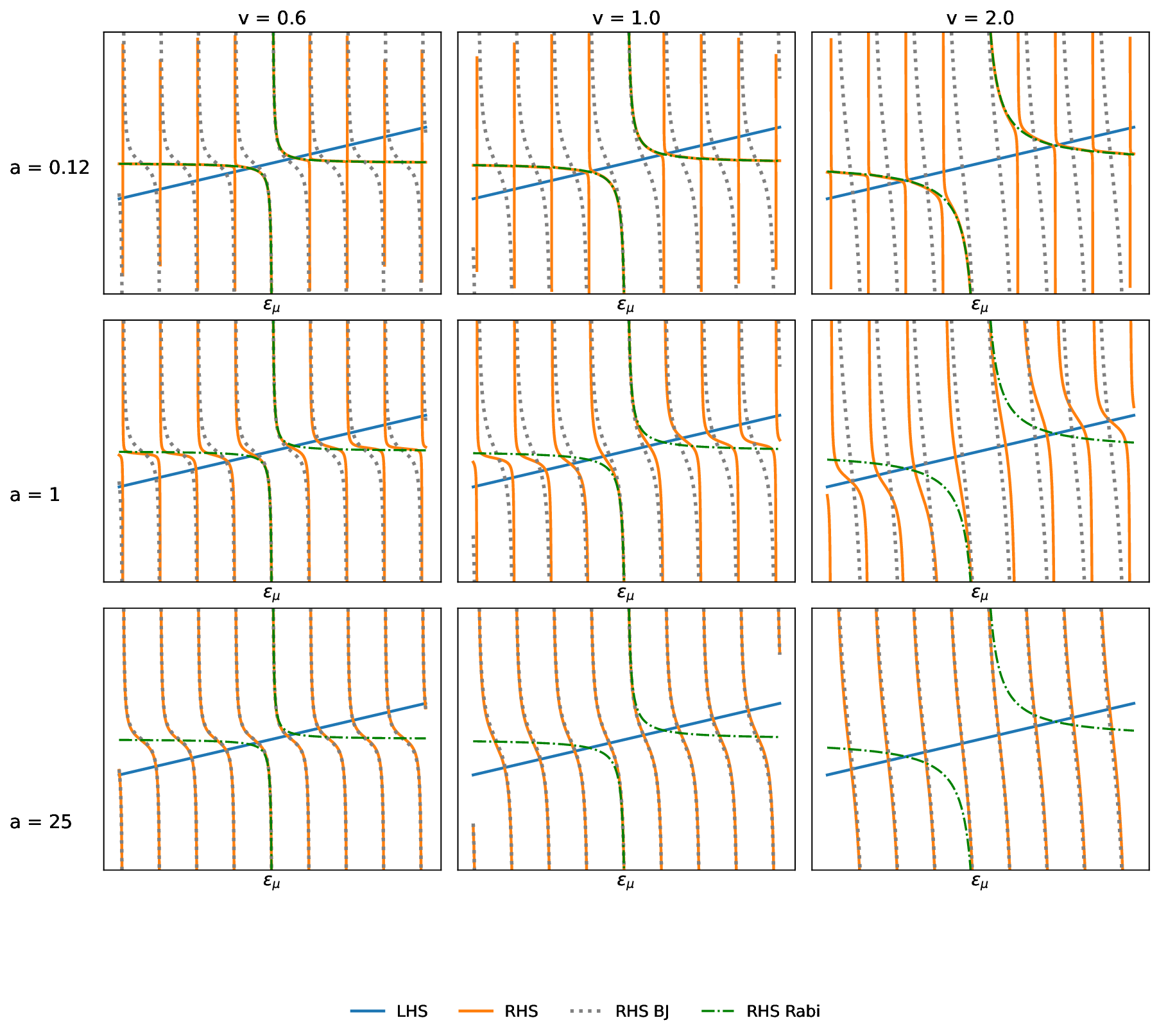}
    \caption{Graphical representation of the transcendental eigenvalue equation \eqref{final-eig} for different choices of the parameters $a$ 
and $v$, keeping $\delta=1$. 
In each plot, the left hand side
of the equation is shown in blue and the
right hand side is shown in orange. The intersections of these two give the eigenvalues. To compare to the limiting behavior, we also show the
RHS of the Bixon-Jortner limit \eqref{BJ-limit} 
in dashed gray, and the Rabi limit \eqref{eq:rabi-limit-eigvals} 
in dot-dashed green. 
The top row of three figures have a small value of the
dimensionless resonance width $a$, hence correspond to the
decoupled Rabi limit described in Section \ref{decoupled-rabi}. %
In this case the eigenvalues are either integers, or the two Rabi values
given in \eqref{eq:rabi-limit-eigvals}. 
The solutions in this case lie on either vertical lines 
or the Rabi curves shown.
The bottom three rows have large values 
of $a$, and thus are well-approximated by the Bixon-Jortner
case. Thus in this case, the eigenvalues are on the Bixon-Jortner
curves.
The middle row has intermediate values of $a$, and has
behavior that is specific to the Lorentzian version of the generalized
Bixon-Jortner system.
In each row, the three plots correspond to three separate values of the overall coupling scale $v$. The case of larger
$v$ corresponds to larger values of $v/\delta$, 
thus approximates the continuum limit.}
    \label{eigval-graphs}
\end{figure}

\section{Dynamics of the discrete state}\label{6}

We next investigate the probability $|\langle\varphi|\psi(t)\rangle|^2$ of 
finding the system in the discrete state at time $t$ if we start with the discrete state at $t=0$: $|\psi(0)\rangle=|\varphi\rangle$. Using the expansion \eqref{eq:eigvec-varphi-component} 
of $|\varphi\rangle$ in terms of the
energy eigenstates, we get the time evolution of $|\psi(t)\rangle$ via
\begin{align}
    |\psi(t)\rangle=\sum_\mu e^{-iE_\mu t}|\psi_\mu\rangle\langle \psi_\mu|\varphi\rangle\,,
\end{align}
where once again we are using units with $\hbar =1$.
The probability amplitude for $|\varphi\rangle$ 
at time $t$ is,
\begin{align}
    \langle \varphi|\psi(t)\rangle &= \sum_\mu  e^{-iE_\mu t}|\langle \varphi|\psi_\mu\rangle|^2 
    \\
    &=\sum_{\mu} \frac{e^{-iE_\mu t}}{1 + 
    \frac{v^2}{\delta^2}S_2(\epsilon_{\mu}, a) }\label{eq:sum-dynamics}
\end{align}
where we used \eqref{eq:eigvec-varphi-component} to get  $|\langle \varphi|\psi_\mu\rangle|^2$.

\subsection{The continuum limit}
Using \eqref{prob-density} we get the probability amplitude $ \langle \varphi|\psi(t)\rangle$ as 
\begin{align}
     \langle \varphi|\psi(t)\rangle=\delta \sum_\mu \frac{W^2\gamma}{\pi} \frac{e^{-iE_\mu t}}{(E_\mu(E_\mu-E_\varphi)-W^2)^2+\gamma^2(E_\mu-E_\varphi)^2}
     \label{eq:time-dep-series}
\end{align}
As $\delta \to 0$, the sum $\delta\sum_\mu f(E_\mu)$ becomes the integral\footnote{For a more detailed justification of 
this limit, see Appendix \ref{appendix:integral-limit}.}
$\int_{-\infty}^\infty f(E)dE$,
\begin{align}
      \langle \varphi|\psi(t)\rangle=\frac{W^2\gamma}{\pi}\int^\infty_{-\infty}dE\frac{e^{-iE t}}{(E(E-E_\varphi)-W^2)^2+\gamma^2(E-E_\varphi)^2}\,.
      \label{eq:time-dep-integral}
\end{align}
Focusing on the case $E_\varphi=0$, we get,
\begin{align}\label{barnett-radmore-integral}
     \langle \varphi|\psi(t)\rangle=\frac{W^2\gamma}{\pi}\int^\infty_{-\infty}dE \frac{e^{-iE t}}{(E^2-W^2)^2+\gamma^2E^2}\,.
\end{align}
As expected, this is precisely the same integral that describes
the time evolution for the coupling of a discrete state to a true continuum with a Lorentzian coupling, 
which is covered in detail in, e.g., 
\cite{barnett2002methods65}---see formula 6.5.45 on page 207. 
The integral \eqref{barnett-radmore-integral} is evaluated for 
$t>0$ by using contour integration, and one gets, 
%
\begin{align}\label{time-evo-inital}
    \langle \varphi|\psi(t)\rangle=W^2\gamma i \left[\frac{e^{-iE_+ t}}{E_+[2(E_+^2 -W^2)+\gamma^2]}+\frac{e^{-iE_- t}}{E_-[2(E_-^2 -W^2)+\gamma^2]}\right]\,,
\end{align}
where,
\begin{align}
     E_\pm=\frac{[-i\gamma\pm\sqrt{(4W^2-\gamma^2)}]}{2}\,.
\end{align}
The probability of finding the system in state $|\varphi\rangle$
is thus (see equation 6.5.47 in \cite{barnett2002methods65}),
\begin{align}\label{prob-of-initial}
    |\langle \varphi|\psi(t)\rangle|^2=W^4\gamma^2\left|\frac{e^{-iE_+ t}}{E_+[2(E_+^2 -W^2)+\gamma^2]}+\frac{e^{-iE_- t}}{E_-[2(E_-^2 -W^2)+\gamma^2]}\right|^2\,.
\end{align}

\subsection{Visualization and commentary}

We next show the dynamics of the generalized
(Lorentzian) BJ system starting in the discrete state using various 
parameter settings, and compare the behavior to known behavior of 
limiting cases. For plotting the limiting cases, we use the analytical formulas for each limit. For plotting the 
dynamics of the full system, we use a numerical approach
where we form a matrix version of the Hamiltonian 
\eqref{eq:system-defn}
using $602\times 602$ matrices (i.e., using $601$ ladder states
going from $n=-300$ to $n=300$, together with the discrete state), obtain the eigenvalues 
and eigenvectors numerically in Python's NumPy package,
and use the formula \eqref{eq:sum-dynamics} to obtain the dynamics.

\begin{figure}[htp]
    \centering
    \includegraphics[width=0.9\textwidth]{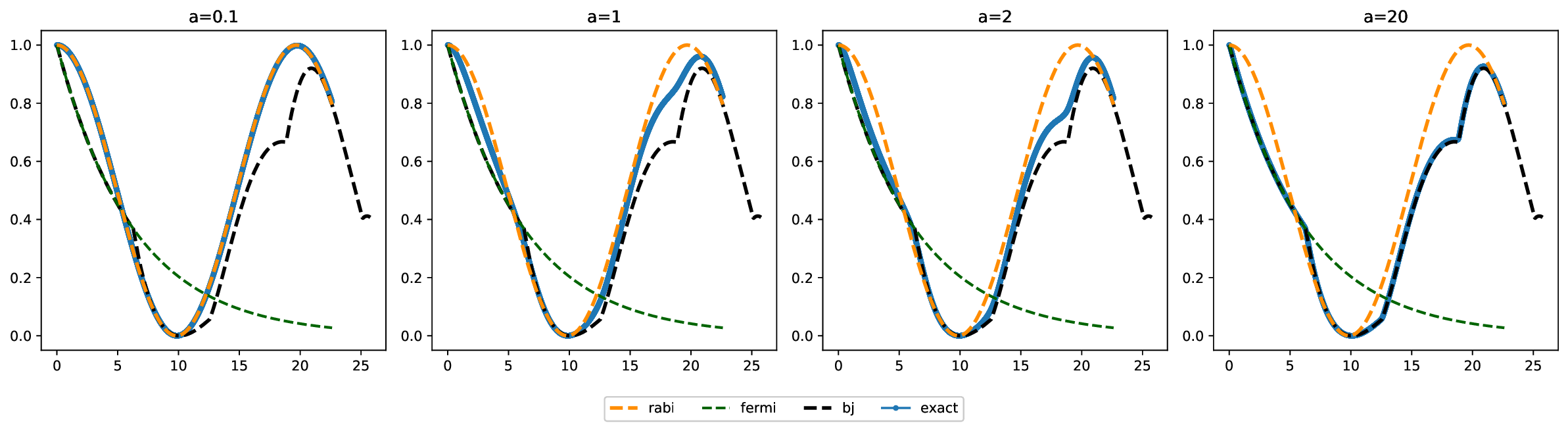}
    \caption{Dynamics of the initial discrete state for different width parameters $a$,
    superposed with pure Rabi oscillations, pure exponential decay,
    and the Bixon-Jortner decay and revivals. In each plot, the solid blue
    curves show the dynamics of the generalized BJ-system, and 
    the dashed curves show the superposed limiting systems.
    These plots correspond
    to the $\beta=0.5$ plot of the BJ system
    in \cite{barnett2002methods}, page 29, 
    which in our notation corresponds to $v\approx 0.16$, $\delta=1$. As $a$ 
    goes from $a=0.1$ to $a=20$, the system behavior interpolates between
    Rabi oscillations and the BJ revival dynamics given in \cite{barnett2002methods}. }
    \label{fig:decay-plots-1}
\end{figure}

\begin{figure}[htp]
    \centering
    \includegraphics[width=0.9\textwidth]{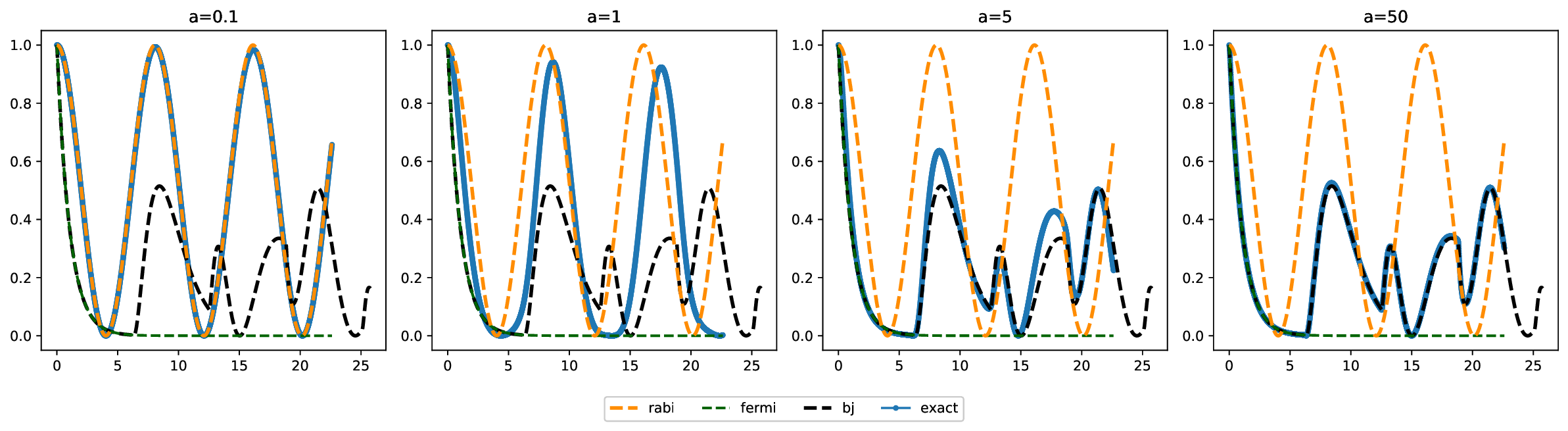}
    \caption{Same as in Figure \ref{fig:decay-plots-1}, this time using the Barnett-Radmore setting of $\beta=3$, which in our notation corresponds to 
    $v\approx 0.39, \delta=1$. Once again, the behavior interpolates between
    Rabi oscillations and the BJ revival dynamics as $a$ is increased.}
    \label{fig:decay-plots-2}
\end{figure}

In Figures \ref{fig:decay-plots-1} and \ref{fig:decay-plots-2} we 
show the probability $|\langle\varphi|\psi(t)\rangle|^2$ as a function of time
for the generalized BJ system,
using parameters settings that correspond to Figures 2.4 and 2.5 in 
\cite{barnett2002methods}.
We superimpose three plots on top of the generalized BJ dynamics to 
show how the generalized system interpolates between 
different types of behavior. Namely, we show, 
\begin{itemize}
    \item The time dependence as given by the exact solution of the BJ 
system
(see Equation 2.5.8 in \cite{barnett2002methods})\footnote{This equation
contains a sign error as we confirmed with Prof. Barnett via email. 
The plots here use a corrected version of the formula.}
\item The Rabi oscillations as given by the corresponding Rabi eigenvalues
\eqref{Rabi-eigenval}
\item An exponential decay whose rate is given by the Fermi golden rule.
\end{itemize}
As can be seen, in both figures, 
the dynamics of the generalized BJ system interpolates
between Rabi oscillations and the pure BJ decay-revival dynamics. The 
sharp transitions that happen at $t=n\delta$ for the BJ system
are smoothed by the generalized BJ system.

\begin{figure}[htp]\label{exp-decay}
    \centering
    \includegraphics[width=0.9\textwidth]{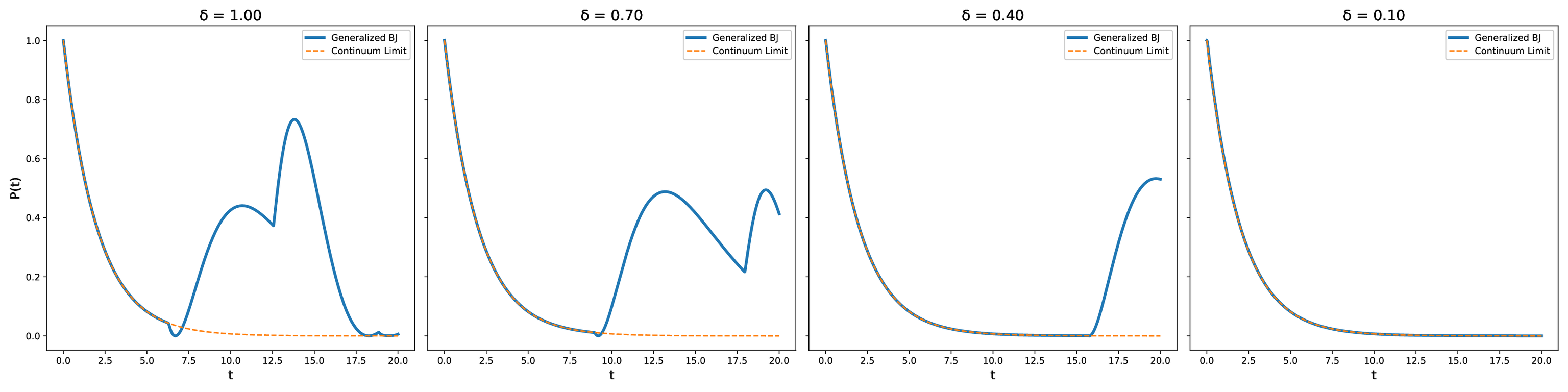}
    \caption{The time dependence of $|\langle\varphi|\psi(t)\rangle|^2$ for an 
    initial discrete state as one approaches the continuum limit of small $\delta$. All plots use the same values for $W$ and $\gamma$, which
    determine the continuum limit through Equation \ref{prob-of-initial}. For this set of plots, we have $W=8.66$, $\Gamma=0.5$, so $\gamma\gg W$, which in the continuum limit gives unoscillatory (overdamped) exponential decay.    
    }
    \label{fig:decay-plots2}
\end{figure}

\begin{figure}[htp]\label{oscillating-decay}
    \centering
    \includegraphics[width=0.9\textwidth]{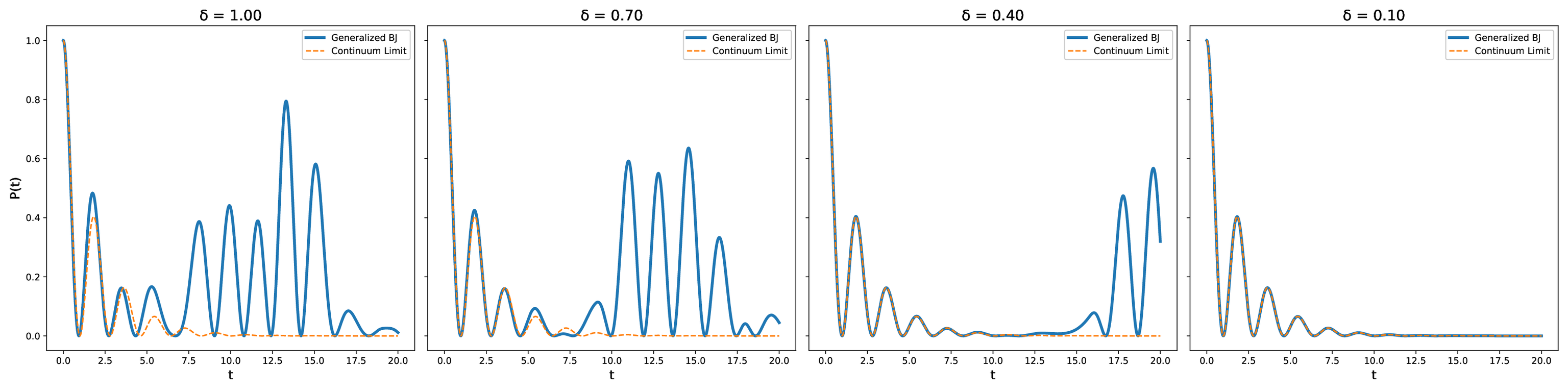}
    \caption{The time dependence of $|\langle\varphi|\psi(t)\rangle|^2$ for an 
    initial discrete state as one approaches the continuum limit of small $\delta$. All plots use the same values for $W$ and $\gamma$, which
    determine the continuum limit through Equation \ref{prob-of-initial}. For this set of plots, we have $W=1.75$, $\Gamma=12.25$, so $\gamma\ll W$, which in the continuum limit gives an oscillatory exponential decay.    
    }
    \label{fig:decay-plots}
\end{figure}

\begin{figure}[htp]
    \centering
    \includegraphics[width=0.9\textwidth]{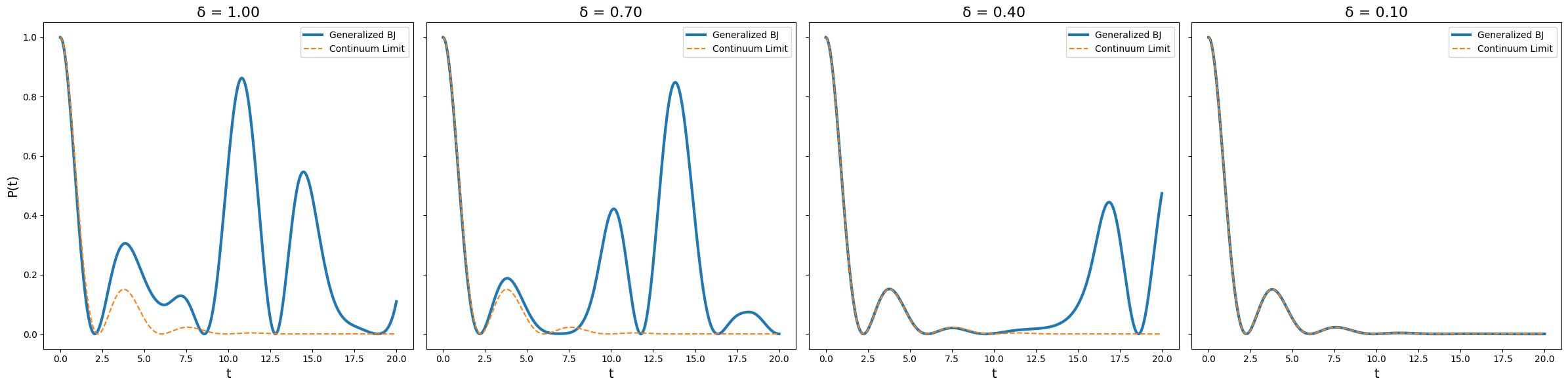}
    \caption{Same as  in Figures \ref{fig:decay-plots2} and \ref{fig:decay-plots}, this time using
    an intermediate ratio between $W$ and $\gamma$. $\Gamma=3, \gamma=0.5, W=0.86, a=(0.5, 0.71, 1,25, 5=, v=(0.69, 0.57, 0.43, 0,21)$}
    \label{fig:decay-plots-intermediate}
\end{figure}

In Figures \ref{fig:decay-plots2}, \ref{fig:decay-plots}, and
\ref{fig:decay-plots-intermediate},
we demonstrate the approach to the continuum limit by choosing progressively 
smaller values of $\delta$, and superpose the known continuum solution
of the dynamics for the continuous Lorentzian coupling from Equation \eqref{prob-of-initial}. 
Since the parameters that describe the continuum behavior are $W=\sqrt{\Gamma 
\gamma/2}$ and $\Gamma$ (see Equation \eqref{time-evo-inital}), in order to investigate 
different types of behavior in the continuum setting 
we plot the dynamics for different regimes in the $W-\Gamma$ space.

As discussed in detail by \cite{tannoudji1992atom}, even in the 
real continuum case, a narrow coupling profile results in approximate Rabi 
oscillations for an initial discrete state. In Figure \ref{Rabi-continuum}, 
we show this approximate behavior by plotting the dynamics in 
a small $\delta$ (approximate continuum) setting
for smaller and smaller values of the coupling
width $\gamma$, showing the reduction in decay, approaching true
Rabi oscillations.

\begin{figure}[htp]\label{Rabi-continuum}
    \centering
    \includegraphics[width=0.9\textwidth]{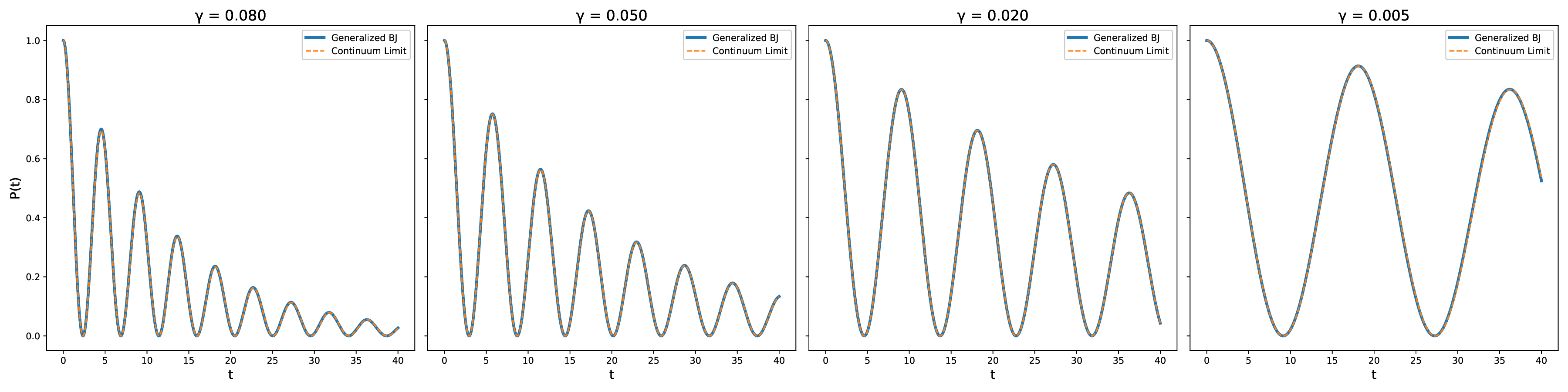}
    \caption{The approach to Rabi oscillations for the generalized BJ system 
    near the continuum limit, which is realized by using a small $\delta$ ($0.005$) for all the plots. As the coupling width $\gamma$ gets small, the decay behavior reduces, making the system approach Rabi oscillations.
    }
    \label{Rabi-continuum}
\end{figure}

\section{Conclusion}\label{conc}
We defined and gave a semi-analytical solution for the diagonalization of 
a model quantum mechanical system that interpolates between some of the 
most fundamental prototypes of quantum mechanical behavior. Defined in terms of
a ladder of states coupled to a separate, discrete state with
a variable coupling, this system approaches the Rabi two-state system in the narrow resonance limit, and the Bixon-Jortner quasi-continuum in the
wide resonance limit, covering oscillatory, decay, and revival dynamics as part of these limits. 
In the quasi-continuum limit of small ladder spacing,
the system gives both the Fermi golden rule exponential 
 decay of the Wigner-Weisskopf model, and more general oscillatory decay 
 in the Fano systems investigated in the quantum optics literature.
 In effect, this system fills in the ``missing square'' of a $2\times 2$
 table of model systems given in Table \ref{tab:coupling_models}.
 
We hope the wide range of behavior covered and the relatively tractable 
analytical formulation will allow this system to be of use to researchers and 
students who depend on model systems to develop understanding and intuition for
more complicated systems. Furthermore, we believe this generalized Bixon-Jortner 
system deserves further studies of its behavior.

\paragraph{Acknowledgement.} This research was supported in part
by Boğaziçi University BAP Program,
Project number 20404, project code 25B03D3.
We would like to thank Prof. Stephen M. Barnett for helpful 
communication regarding the dynamics of the Bixon-Jortner discrete
state.

\appendix

\section{Justification of the Integral Limit}
\label{appendix:integral-limit}

During our computation of the time-dependent probability amplitude
of the discrete state, we replaced the sum \eqref{eq:time-dep-series} with the integral \eqref{eq:time-dep-integral} in 
the continuum limit of $\delta\to 0$. While this is a reasonable
procedure, one could imagine pathological behavior in the 
$\delta\to 0$ limit where the number of eigenvalues in the interval $[n\delta, (n+1)\delta]$ depends 
on $\delta$ in a complicated way.
In this section we show 
that this pathology does not occur for the generalized BJ system,
in the sense that in the $\delta\to 0$ limit, there is 
only a single eigenvalue in each such $\delta$-interval. 

Let us look at the transcendental equation giving the dimensionless eigenvalues, 
\begin{equation}
\epsilon_{\mu} = \epsilon_{\varphi} + \frac{\pi v^2}{\delta^2} \left[
\frac{ \cot(\pi \epsilon_{\mu}) +  \alpha(a) \epsilon_{\mu} }{1 + 
(\epsilon_{\mu}/a)^2} \right]\,.
\end{equation}
The dimensionful version written in terms of the
constants $\gamma$ and $\Gamma$ can be written as
\begin{equation}
 g(E_\mu): = E_{\varphi} + \frac{\Gamma}{2} \left[
\frac{ \cot(\frac{\pi}{\delta} E_{\mu}) +  \frac{\coth{\pi a}}{\gamma} E_{\mu} }{1 + 
(E_{\mu}/\gamma)^2} \right]-E_{\mu} =0\,.
\end{equation}
Consider the behavior of the function $g(E_{\mu})$ in the interval $[n\delta, (n+1)\delta]$:  $g(E_{\mu})\to+\infty$ as $E_{\mu}\to n\delta$ from above, and $g(E_{\mu})\to-\infty$ as  $E_{\mu}\to (n+1)\delta$ from below. Thus if we can show that 
$g(E_{\mu})$ is strictly decreasing in this interval, it will
follow that it has a single root in this interval. Now, $E_\varphi$ is constant and $-E_\mu$ is strictly decreasing, thus we only need to show that
\begin{align}
    \frac{ \cot(\frac{\pi}{\delta} E_{\mu}) +  \frac{\coth{\pi a}}{\gamma} E_{\mu} }{1 + 
(E_{\mu}/\gamma)^2}
\end{align}
is strictly decreasing. But since $\frac{1}{1+(E_\mu/\gamma)^2}$ is also decreasing, if we can show that the function
\begin{align}
  f(E_\mu)=  \cot(\frac{\pi}{\delta} E_{\mu}) +  \frac{\coth{\pi a}}{\gamma} E_{\mu}
\end{align}
is strictly decreasing, we are done. Now, the derivative 
\begin{align}
    f'(E_\mu)=-\frac{\pi}{\delta}\frac{1}{\sin^2(\frac{\pi}{\delta}E_\mu)}+  \frac{\coth{\pi a}}{\gamma},
\end{align}
has a maximum in our interval
at $E_{\mu} = (n + 1/2)\delta$, therefore if we can show that this maximum is negative, this will imply that $f(E_\mu)$
is strictly decreasing. This maximum value becomes zero at
$a=a_0$, where $a_0$ is a solution of the equation
\begin{align}
    -\frac{\pi}{\delta}+  \frac{\coth{\pi a_0}}{\gamma}=0\,,
\end{align}
i.e., $\coth{\pi a_0}=\pi a_0$. Since $\coth$ is monotonically
decreasing for positive values of its argument, $f'(E_\mu)$ 
would be guaranteed to be negative if $a> a_0$. Now, we
take the limit $\delta\to 0$ while keeping $\gamma = a\delta$ constant, so for small enough values of $\delta$, $a<a_0$ is guaranteed. In other words, for small enough $\delta$, $f'(E_\mu)<0$ in our interval and therefore $g(E_\mu)$ has a single root 
in $[n\delta, (n+1)\delta]$.


\bibliographystyle{plain}
\bibliography{ref}

@book{allen2012optical,
  title={Optical resonance and two-level atoms},
  author={Allen, Leslie and Eberly, Joseph H},
  year={2012},
  publisher={Courier Corporation}
}

@incollection{cohen1986quantum,
  title={Quantum Mechanics},
  author={Cohen-Tannoudji, Claude and Diu, Bernard and Laloe, Frank},
  publisher={New York, NY (United States); John Wiley and Sons Inc.},
  volume={2},
  edition={2nd},
  chapter={\MakeUppercase{\romannumeral 13}--C--3--b},
}

@book{barnett2002methods,
  title={Methods in theoretical quantum optics},
  author={Barnett, Stephen and Radmore, Paul M},
  volume={15},
  year={2002},
  publisher={Oxford University Press}
}

@article{bixon1968intramolecular,
  title={Intramolecular radiationless transitions},
  author={Bixon, Mordechai and Jortner, Joshua},
  journal={The Journal of chemical physics},
  volume={48},
  number={2},
  pages={715--726},
  year={1968},
  publisher={American Institute of Physics}
}

@article{weisskopf1930berechnung,
  title={Berechnung der nat{\"u}rlichen linienbreite auf grund der diracschen lichttheorie},
  author={Weisskopf, Victor and Wigner, Eugene},
  journal={Zeitschrift f{\"u}r Physik},
  volume={63},
  number={1},
  pages={54--73},
  year={1930},
  publisher={Springer}
}

@incollection{barnett2002methods65,
  title={Methods in theoretical quantum optics},
  author={Barnett, Stephen and Radmore, Paul M},
  volume={15},
  year={2002},
  publisher={Oxford University Press},
  chapter={6.5},
}

@article{fano1961effects,
  title={Effects of configuration interaction on intensities and phase shifts},
  author={Fano, Ugo},
  journal={Physical review},
  volume={124},
  number={6},
  pages={1866},
  year={1961},
  publisher={APS}
}

@article{tannoudji1992atom,
  title={Atom-photon interactions},
  author={Tannoudji, Claude Cohen and Grynberg, Gilbert and Dupont-Roe, J},
  year={1992},
  publisher={New York, NY (United States); John Wiley and Sons Inc.}
}

@incollection{cohen1986quantum1,
  title={Quantum mechanics},
  author={Cohen-Tannoudji, Claude and Diu, Bernard and Laloe, Frank},
  publisher={New York, NY (United States); John Wiley and Sons Inc.},
  volume={1},
  pages={414},
  edition={2nd}
}

@article{tannoudji1992atom2,
  title={Atom-photon interactions},
  author={Tannoudji, Claude Cohen and Grynberg, Gilbert and Dupont-Roe, J},
  year={1992},
  publisher={New York, NY (United States); John Wiley and Sons Inc.},
  pages={53},
}

@article{bluhm1995evolution,
  title={The evolution and revival structure of localized quantum wave packets},
  author={Bluhm, Robert and Kostelecky, Alan and Porter, James},
  journal={arXiv preprint quant-ph/9510029},
  year={1995}
}

@article{narozhny1981coherence,
  title={Coherence versus incoherence: Collapse and revival in a simple quantum model},
  author={Narozhny, NB and Sanchez-Mondragon, JJ and Eberly, JH},
  journal={Physical Review A},
  volume={23},
  number={1},
  pages={236},
  year={1981},
  publisher={APS}
}

@book{cartan1995elementary,
  title={Elementary theory of analytic functions of one or several complex variables},
  author={Cartan, Henri},
  year={1995},
  publisher={Courier Corporation}
}

\end{document}